# Black-Box Assessment of Optical Spectrum Services

**Kaida Kaeval[1,2], Jörg-Peter Elbers[1], Klaus Grobe[1], Marko Tikas[3], Tobias Fehenberger[1], Helmut Grieβer[1], Gert Jervan[2]**

*[1] ADVA Optical Networking SE, Martinsried, Germany*
*[2] Tallinn University of Technology, Tallinn, Estonia*
*[3] Tele2 Estonia, Tallinn, Estonia*
*kkaeval@adva.com*

**Abstract:** A spectral sweep process is introduced to discover performance issues in optical spectrum services. We detect filtering penalty, spectral ripple/tilt and channel crosstalk in field measurements, potentially leading to increased service robustness in low-margin networks. 

## 1. Introduction

Recently, optical spectrum is emerging as new service paradigm in high capacity optical DWDM networks [1]. Relying on open line systems (OLS), network operators can provide optical spectrum to customers who run traffic over it using their own DWDM terminals. To achieve the highest transmission capacity for a spectral slot at hand, customers need to choose the best transponder configuration that satisfies the minimum margin requirements [2]. While transmission impairments make an accurate performance evaluation challenging already for fixed-grid wavelength services [3], a flex-grid Optical Spectrum as a Service (OSaaS) scenario introduces even more variables.

Pre-characterized coherent transceivers have been proposed to estimate the Generalized Signal to Noise Ratio (GSNR) of optical lightpaths [4] and have been proven to reliably predict the Quality of Transmission (QoT) [5]. In this work, we adapt the frequency sweep procedure proposed in [6] and combine it with the Extended Channel Probing (ECP) method [7] to assess the performance of optical spectrum services in a black-box-scenario. We apply this approach to investigate the impact of narrow optical filtering, carrier frequency mismatch, spectral ripple/tilt, and adjacent channel crosstalk and discuss how to define spectrum services in a robust way.

## 2. System description and measurement set-up

Following the definitions in ITU-T Recommendation G.807 [8], we define an optical spectrum service as an optical media channel (MC) representing a transparent lightpath through an optical network on a predetermined frequency slot. A media channel supports zero or more optical tributary signals (OTSi) each of which may be carried in its own Network Media Channel (NMC). The MCs and NMCs can use guard bands to limit crosstalk between adjacent OTSi in and across adjacent MCs.

In real life scenarios, it is difficult to evaluate the media channel mask or channel transfer function determined by the cascade of optical filters along the lightpath. While ASE noise loading or attenuation profile measurements can determine the exact mask characteristics, they do not capture a possible GSNR variation over the media channel bandwidth.

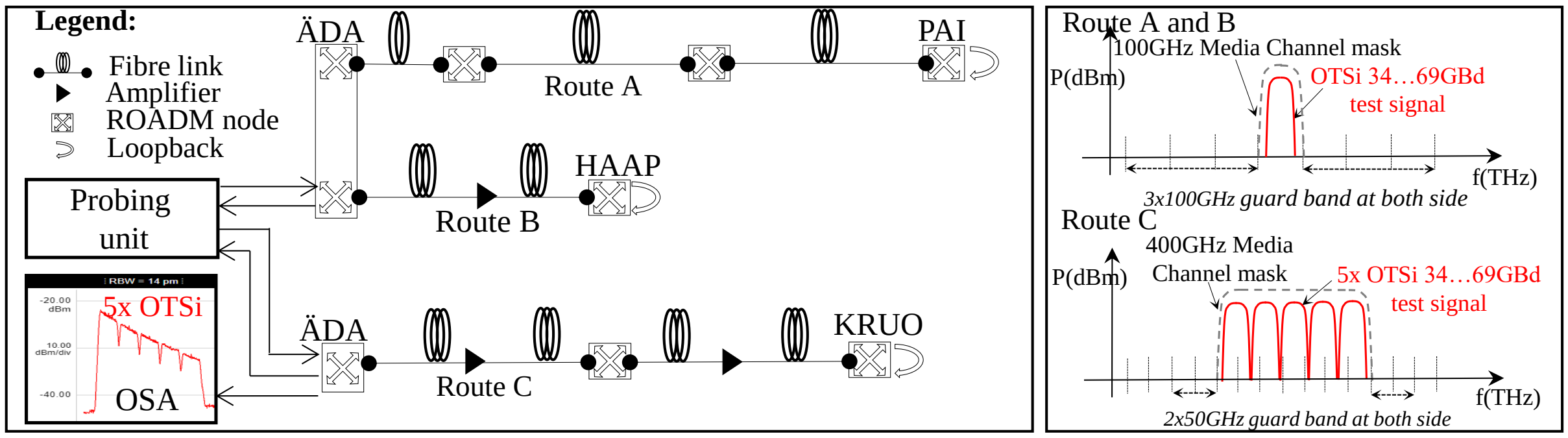

| Route | Network | DCM data | Looped link length | No of spans | Central frequency | Nominal slot width | Tested service width |
|---|---|---|---|---|---|---|---|
| A | Regional | DCF only | 241km | 6 | 192.3THz | 100GHz | 100GHz |
| B | Regional | DCF+DCG Mix | 284km | 4 | 192.3THz | 100GHz | 100GHz |
| C | Long-haul | DCM free | 1792 km | 24 | 193.95THz | 50GHz | 400GHz |

Fig. 1. General test set-up, spectral assignments and route data

To assess the feasibility of OSaaS offerings in field deployments, we tested optical spectrum services using MCs over three different routes in Tele2 Estonia's ROADM-based production network. The test set-up and available parameters of route A, B, and C are illustrated in Fig.1. To extend the transmission distance and to allow single-ended measurements in ÄDA, the spectrum services were looped back in the far end ROADMs.

The spectrum services on route A and B run over 10 Gb/s legacy DWDM links with fixed 100 GHz channel grid. Route A uses dispersion compensating fibers (DCFs) for optical dispersion compensation, whereas route B uses a mix of DCFs and dispersion compensating gratings (DCGs). In the terminal ROADMs in ÄDA, the optical user spectrum was added and dropped by arrayed waveguide grating (AWG) filters.

Route C runs over a dispersion-uncompensated long-haul DWDM link with flex-grid ROADMs optimized for 100 Gb/s transmission. Here, the optical user spectrum was added and dropped at a free terminal ROADM port using 8:1 splitter/combiner module in ÄDA without applying additional optical filtering.

A 100 GHz narrow-band spectrum service was investigated on routes A and B, whereas a 400 GHz wide-band service was used on route C. The channel performance on all routes was tested with 200 Gbit/s 69 GBd DP-QPSK, 200 Gbit/s 46 GBd DP-P-16QAM, and 200 Gbit/s 34 GBd DP-16QAM probing signals with root raised cosine (RRC) spectral shape (roll-off factor r = 0.19). To obtain comparable results, we maintain a constant ratio of signal power and symbol rate and adjust the signal power at each symbol rate accordingly when switching modulation formats. This is possible when the DWDM line systems and their amplifiers operate in constant gain mode. The system performance is evaluated by a generalized SNR (GSNR) which is derived from the measured Q-factor and normalized to the symbol rate [7]. By sweeping in 6.25 GHz steps over the allocated spectrum, we investigate the impact of filtering, frequency misalignment, spectral ripple/tilt effects and adjacent channel crosstalk from the perspective of an optical spectrum user.

## 3. Results and interpretation

Fig. 2 shows the results of the channel probing on the three different routes. While the nominal MC width on route A and B is 100 GHz, the large predicted GSNR difference between the three modulation formats reveals that the usable effective optical bandwidth must be much smaller. Given that a 6.25GHz mismatch of a 34GBd signal already causes a noticeable penalty, we can conclude that the effective optical channel bandwidth is lower than 50GHz. In Fig. 2a, up to 2dB degradation in the estimated GSNR is observed for a 200Gbit/s 34GBd DP-16QAM signal configuration in case of 18.75GHz offset. With the same offset, 200Gbit/s 46GBd DP-P-16QAM and 200Gbit/s 69GBd DP-QPSK signals are already experiencing outage as the attainable GSNR is insufficient.

The results in Fig. 2b also indicate a misalignment of the nominal center frequency on route B in addition to severe filtering penalty for the 200Gbit/s 69GBd DP-QPSK signal format. Without sweeping, this misalignment could have been left undiscovered and already a small fluctuation in the transceiver wavelength could cause a severe degradation of the service quality, especially with 46GBd DP-P-16QAM modulation, that generally requires higher margin and is more prone to filtering penalty, than a 200Gbit/s 69GBd DP-QPSK signal. To increase the robustness, the transceiver frequency should be fine-tuned to minimize frequency misalignment and to obtain the best media channel performance.

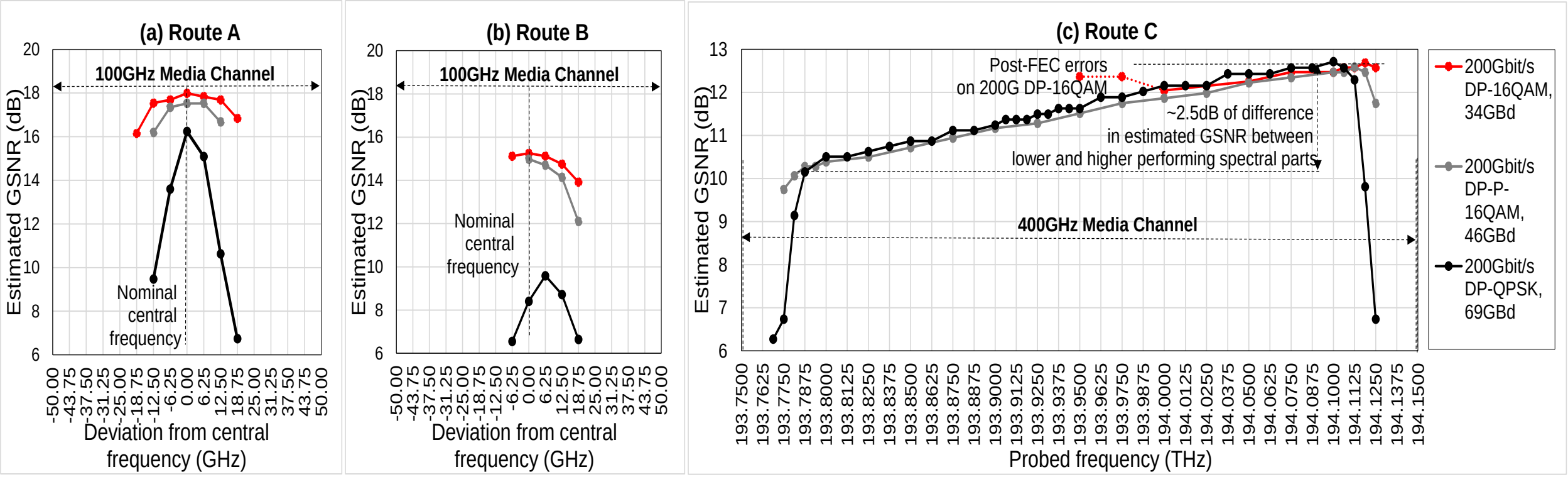

Fig. 2 Channel probing sweep results on a) route A, b) route B and c) route C

Fig.2c presents the GSNR variation over the 400GHz spectrum slot on route C. While all three modulation formats predict a similar GSNR performance, there is a 2.5 dB GSNR tilt over the media channel width and DP-16QAM (red line with extended dots) lacks sufficient margin to work at the lower 2/3 part of the slot. The knowledge of the GSNR shape can help to maximize the total throughput of the spectral slot. Based on estimated GSNR and our example route, 300 Gbit/s 69GBd DP-P-16QAM modulation could be used at the lower end of the spectrum, while 300 Gbit/s 52 GBd DP-16QAM signals are possible at the higher end. The obtained GSNR information could also be used for a power pre-emphasis at the transmitter to equalize the signal performance across the spectral slot.

In colorless splitter/combiner based access architectures, there is a threat of impacting the neighboring channels, when signals are operated too close to the spectral slot borders and no guard-bands are used. Degradation is caused by power leakage into the neighboring slot. To investigate the severity of this scenario, we reconfigured the single 400 GHz media channel to five independent 75 GHz slots with one optical carrier per each slot. We then swept the central carrier frequency from one edge of the middle slot to the other edge.

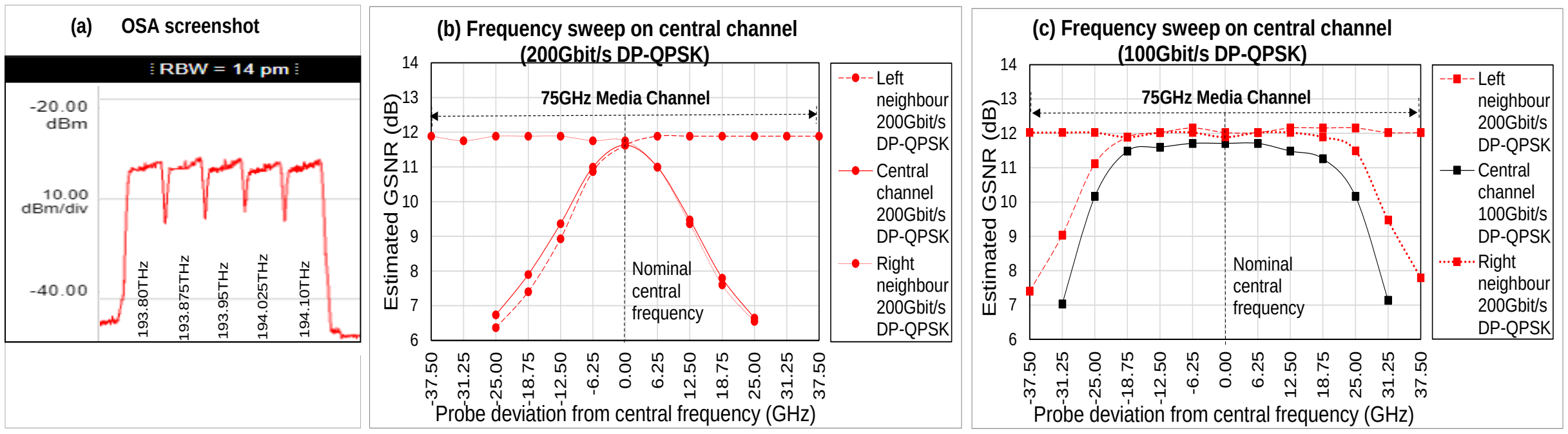

*Fig. 3. Sweeping impact on the neighbouring channels in colourless directionless architectures*

Fig. 3a illustrates the spectral arrangement of the signals for a crosstalk test with five 69GBd 200G DP-QPSK channels. Crosstalk is only noticeable between nearest neighbor channels. Fig. 3b shows that channels with equal symbol rates experience a similar penalty from decreased carrier spacing. This is understandable by the spectral symmetry. The spectral width of each signal is determined by its symbol rate (SR) and the roll-off factor r as (1+r)*SR and can be used to estimate how far channels need to be spaced apart to minimize crosstalk. A reduction of 6.25 GHz, 12.5 GHz and 18.75 GHz in channel spacing results in a degradation of the estimated GSNR by 0.8dB, 2.5dB and >4.4dB, respectively. Such degradation may not be acceptable in low-margin networks. In Fig. 3c, all side channels are maintained in 200Gbit/s 69GBd DP-QPSK format, while the central channel is reconfigured to 100G DP-QPSK modulation (solid black line). Due to the different symbol rate of the central channel, a spectral overlap with neighboring channels only happens at larger frequency offsets. At the same offset, the higher symbol-rate channels experience less crosstalk than the central channel due to their larger spectral width. Crosstalk becomes noticeable at 25GHz offset from the nominal central frequency of the test channel, where degradation in estimated GSNR is 1.2dB for 100Gbit/s 34GBd DP-QPSK channel and 0.9dB for 200Gbit/s 69GBd DP-QPSK neighbor. From our test results, it is recommended to run more robust signal configurations at the edges of a spectral slot to account for possible power or central frequency fluctuations in neighboring channels. Operators should also consider to allocate guard bands between media channels of different customers taking the maximum spectral width of transmitted signals into account.

## 4. Conclusions

We tested multiple Optical Spectrum as a Service (OSaaS) scenarios using a combined sweep and extended channel probing method. We verified that both parts of the procedure – frequency sweeping and symbol rate variable probing - do have their unique benefits in pin-pointing configuration issues and simplifying performance estimations. Based on this data, OSaaS users can take fully informed decisions, on how to leverage their spectral resources in the most efficient way without prior knowledge on the media channel envelope or impacting neighboring channels.

## 5. References

[1] ADVA, (2020), "ADVA Unlocks Network Potential With Spectrum As A Service ", press release, 12/03/2020
[2] Y. Pointurier, "Design of low-margin optical networks," JOCN, vol. 9, no. 1, pp. A9-A17, Jan. 2017
[3] A. P. Vela et al., "BER Degradation Detection and Failure Identification in Elastic Optical Networks," JLT, vol. 35, no. 21, Nov. 2017
[4] E. Rivera-Hartling et al., "Design, Acceptance and Capacity of Subsea Open Cables," in JLT, vol. 39, no. 3, Feb. 2021
[5] A. Ferrari et al., "GNPy: an open source application for physical layer aware open optical networks," JOCN, vol. 12, no. 6, 2020
[6] A. C. Meseguer et al., "Automated Full C-Band Technique for Fast Characterization of Subsea Open Cable G-SNR," in Proc. ACP2019, pp. 1-3, Chengdu, China
[7] K. Kaeval et al., "Channel Performance Estimations with Extended Channel Probing," in Proc. ITG-PN 2020, pp. 1-5, online.
[8] ITU-T Rec. G.807, online https://www.itu.int/rec/T-REC-G.807/en